\documentclass[twocolumn,showpacs,preprintnumbers,amsmath,amssymb]{revtex4}

\usepackage{graphicx}

\newcommand{\fHe}{\ensuremath{{}^{4}\!\mathrm{He}}}
\newcommand{\aRb}{\ensuremath{{}^{87}\!\mathrm{Rb}}}
\newcommand{\invs}{\ensuremath{\mathrm{s}^{-1}}}
\newcommand{\Done}{\ensuremath{\mathrm{D}_{1}}}

\newcommand{\alignment}{\mathcal{A}_{0}}

\newcommand{\Hz}{\ensuremath{\mathrm{Hz}}}

\newcommand{\grscale}{0.48}

\newcommand{\myvector}{\boldsymbol}

\newcommand{\liaphase}{\ensuremath{\phi_l}}
\newcommand{\dphase}{\ensuremath{\Delta\alpha}}
\newcommand{\sGamma}[1]{\ensuremath{\Gamma_{\kern-0.23em #1}}}

\begin{document}

\title{Experimental study of laser detected magnetic resonance based on atomic alignment}

\author{Gianni Di~Domenico}
\email{gianni.didomenico@unifr.ch}
\author{Georg~Bison}
\author{Stephan~Groeger}
\author{Paul~Knowles}
\author{Anatoly~S.~Pazgalev}
\author{Martin~Rebetez}
\author{Herv\'e~Saudan}
\author{Antoine~Weis}

\affiliation{%
Physics Department, University of Fribourg, Chemin du Mus\'ee 3,
1700 Fribourg, Switzerland
}%

\date{September 20, 2006}

\begin{abstract}
We present an experimental study of the spectra produced by
optical/radio-frequency double resonance in which resonant
\emph{linearly} polarized laser light is used in the optical pumping
and detection processes. We show that the experimental spectra
obtained for cesium are in excellent agreement with a very general
theoretical model developed in our group~\cite{DRAMtheo} and we
investigate the limitations of this model. Finally, the results are
discussed in view of their use in the study of relaxation processes
in aligned alkali vapors.
\end{abstract}


\maketitle

\section{Introduction}
\label{sec:intro}

%
Optically pumped alkali vapor magnetometers (OPM) using circularly
polarized light are widely used in many applications.  One of the
most common realizations is the so-called $M_x$ magnetometer
\cite{Blo62}, where circularly polarized light is used to create a
macroscopic spin polarization (orientation) in a vapor of alkali
atoms at an angle of $45\deg$ with respect to the static magnetic
field $\myvector{B}_{0}$ under investigation. A small additional
magnetic field $\myvector{B}_{1}(t)$, hereafter referred to as
radio-frequency or rf~field, oscillating at a frequency close to the
Larmor frequency, induces oscillations of sublevel coherences in the
sample.  These coherent oscillations lead to a modulation of the
transmitted light power at the rf~frequency, which shows a resonant
behavior when the rf frequency is close to the Larmor frequency.
Either the resonant amplitude change or the resonant phase change
can be used to infer the magnetic field modulus. Discharge lamps are
commonly used light sources for alkali OPMs. Such lamps emit a broad
spectrum and can be used for optical pumping only when a single fine
structure component, typically the \Done-line for alkalis, is
isolated by means of an appropriate filter \cite{Happer72}.

%
Pumping with circularly polarized light builds up a vector
polarization (orientation), while pumping with linearly polarized
light leads to the creation of a tensor polarization (alignment) in
the atomic ground state, provided that it has an angular momentum
larger than $1/2$.
The ground state of alkali atoms has an electronic angular momentum
$J=1/2$, which can not be aligned. However, the hyperfine
interaction with the nuclear spin $I$ splits the ground state into
two hyperfine levels with total angular momenta $F_\pm=|I\pm 1/2|$,
which can be aligned provided $F\geq 1$. An alignment can therefore
be prepared, or detected, if the light source has a sufficient
spectral resolution to excite a single hyperfine transition. In
general, this is not the case for the Doppler and
pressure--broadened spectrum of discharge lamps.

%
The first detection of an optically induced alignment was reported
in \fHe\cite{BellBloom61}. Later, both DC signals and signals
oscillating at the fundamental and at the second harmonic of the
alignment precession frequency have been observed in \aRb{} via
birefringence measurements \cite{Happer1970} using a double beam
technique.  In \cite{AlexAlign1990} the authors observed the
alignment in a lamp-pumped \aRb{} vapor by analyzing the induced
linear birefringence of the vapor while using the same light beam
for pumping and probing the atoms. The creation and detection of
atomic alignment with radiation from a discharge lamp becomes
possible when using an isotope filtering technique
\cite{AlexAlign1990,Happer1970} for isolating specific hyperfine
components. However, that technique is restricted to Rb and can
neither be applied to Na nor to K, where the lamp emission spectrum
leaves the hyperfine structure unresolved, nor to Cs for which no
isotope filters are available.

%
Due to their narrow spectral widths (several MHz compared to the
Doppler broadening of several hundred MHz), tunable lasers provide a
way to selectively drive an isolated optical hyperfine transition, a
very effective way to create an alignment with linearly polarized
light.  Narrow-band tunable laser sources are therefore an
interesting alternative to spectral lamps for magnetometric
applications.  Their use in $M_x$ magnetometers has been
demonstrated \cite{aleMx} and it could be shown that laser-based
alkali magnetometers yield a superior magnetometric performance
compared to lamp pumped devices \cite{aleMx,CompLmpLas}.
A~discussion of laser pumping with circularly and linearly polarized
light in \fHe{} can be found in
\cite{Gilles1992,Gilles_He4LsMag_2001}.  Those authors investigated
several magnetometry techniques using both orientation and alignment
signals and observed magnetic resonances by applying either
rf~fields, light intensity modulation, polarization modulation, or
laser frequency modulation.  A~variant of the latter technique in
which the polarization of the transmitted light was measured, rather
than its intensity, was realized with \aRb{} by
\cite{Budker2000,BudkerPRA2002}.

%
In a recent publication~\cite{DRAMtheo}, our group presented a
theoretical study of the spectra produced by optical/radio-frequency
double resonance devices in which resonant \emph{linearly} polarized
light is used in the optical pumping process. That study led to very
general algebraic results, valid for atomic states with arbitrary
angular momenta, for arbitrary rf intensities, and for arbitrary
geometries, valid however only for low laser light power. Therefore,
our goal was to investigate experimentally the domain of validity of
this theoretical model.

In this work, we present an experimental investigation of the
magnetic resonance spectra produced in a cesium vapor in which an
alignment is created and detected by a single linearly polarized
laser beam. We show that the experimental spectra obtained for
cesium are in excellent agreement with the theoretical model
presented in~\cite{DRAMtheo} and we investigate the limitations of
this model. In contrast to prior work reported in
\cite{Happer1970,AlexAlign1990,Gilles1992} we detect the signal at
both the first and second-harmonic of the applied rf field.  This
offers the possibility of detecting the alignment in low magnetic
fields where the quadratic Zeeman splitting is negligible. The
technique reported in \cite{AlexAlign1990} is insensitive to
alignment in small fields since the signals at the first harmonic
cancel each other in the low field limit.

\section{Experiments}
\label{sec:Experiments}

\subsection{Principle}
\label{subsec:principle}

\begin{figure}[b]
\includegraphics[width=\grscale\textwidth, clip]{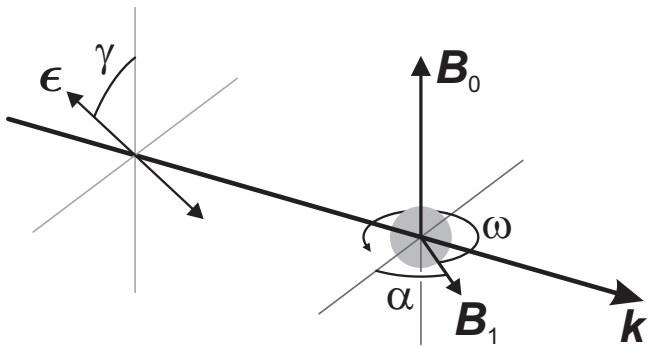}
\caption{Geometry of double resonance spectroscopy involving
linearly polarized light. The oscillating field $\myvector{B}_{1}$
rotates in a plane perpendicular to the static field
$\myvector{B}_{0}$.  The linear polarization vector
$\myvector{\epsilon}$ makes an angle $\gamma$ with the static field
$\myvector{B}_{0}$, and the phase of $\myvector{B}_{1}$ is
characterized by $\alpha$.} \label{fig:geometry}
\end{figure}

The experimental geometry, shown in Fig.~\ref{fig:geometry}, is
identical to the one described in~\cite{DRAMtheo}.  A linearly
polarized laser beam, resonant with a given hyperfine transition,
creates an atomic alignment in a room temperature vapor of cesium
atoms by optical pumping.  This alignment is driven coherently by
simultaneous interactions with the optical field, a static magnetic
field $\myvector{B}_{0}$, a much weaker rotating magnetic field
$\myvector{B}_{1}$, and is affected by relaxation processes.  The
rf~field $\myvector{B}_{1}$ rotates in a plane perpendicular to
$\myvector{B}_{0}$ at frequency $\omega$. The steady state dynamics
are probed by monitoring the transmitted light power.

The precession of the alignment driven by the rf field leads to
modulations of the transmitted power at both the fundamental and the
second harmonic of the rf frequency $\omega$.  Lock-in detection is
used to record spectra of the first and second harmonic signals and
the results are compared with the theoretical predictions
of~\cite{DRAMtheo}. We investigated the effects of the experimental
parameters on the spectral line shapes of the in-phase and
quadrature components of both signals. In particular, we studied the
dependence of the signals on the angle $\gamma$ between the linear
polarization vector and the static field, the dependence on rf
power, and the dependence on light power.

\subsection{Experimental setup}
\label{subsec:expsetup}

\begin{figure}
\includegraphics[angle=270,width=\grscale\textwidth,clip]{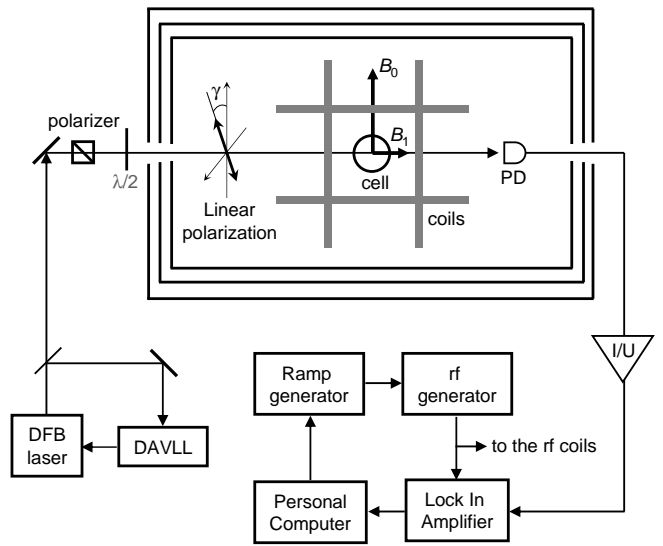}
\caption{Sketch of the experimental setup.  A paraffin-coated glass
cell containing Cs vapor was placed inside a three-layer mumetal
shield. The shield contained Helmholtz coils to produce the static
magnetic field $\myvector{B}_{0}$, and radio-frequency coils to
produce the rotating magnetic field $\myvector{B}_{1}$. The DFB
diode laser was stabilized to the $F_g=4 \rightarrow F_e=3$
transition of the Cs \Done{} line by means of a dichroic atomic
vapor laser lock (DAVLL). A~linear polarizer followed by a half-wave
plate ($\lambda /2$) were used to prepare linearly polarized light
at an angle $\gamma$ with respect to $\myvector{B}_{0}$. The
transmitted light power after the cell was monitored by a
nonmagnetic photodiode (PD) whose signal was amplified by a
low-noise transimpedance amplifier and analyzed by a lock-in
amplifier. The function generator driving the rf field also served
as reference for the lock-in amplifier. The resonance spectra were
recorded by a personal computer which controlled the rf frequency
scan.} \label{fig:expsetup}
\end{figure}

The measurements were done using a room temperature cesium vapor in
vacuum confined within a paraffin-coated spherical glass cell (28~mm
diameter). The experimental setup is shown in
Fig.~\ref{fig:expsetup}.
The cesium vapor cell was isolated from
ambient magnetic fields by a three-layer mumetal shield.  Inside the
shield, a pair of Helmholtz coils produced a static magnetic field
$\myvector{B}_{0}$ of about $3\:\mu\mathrm{T}$ perpendicular to the
light propagation direction.
Additional orthogonal pairs of Helmholtz coils (only one pair is
shown in Fig.~\ref{fig:expsetup}) were used to compensate residual
transverse fields and gradients.
An rf magnetic field $\myvector{B}_{1}$, rotating at approximately
10~kHz in the plane perpendicular to the static magnetic field, was
created by a set of two pairs of Helmholtz coils, wound on the same
supports as the static field coils.
%
%
The laser beam used to pump and probe the atomic vapor was generated
by a DFB diode laser ($\approx 894\:\mbox{nm}$) stabilized to the
$6S_{1/2}, F_g\!=\!4\!\rightarrow\!6P_{1/2}, F_e\!=\!3$ hyperfine
component by means of a dichroic atomic vapor laser lock (DAVLL)
\cite{WiemanDAVLL}.  A~linear polarizer followed by a half-wave
plate prepared linearly polarized light of adjustable orientation
$\gamma$ with respect to $\myvector{B}_{0}$.  After the half-wave
plate, the remaining degree of circular polarization was smaller
than 1\%. The light power transmitted through the cell was detected
by a nonmagnetic photodiode, whose photocurrent was amplified by a
low-noise transimpedance amplifier. The resulting signal was
analyzed by a lock-in amplifier tuned either to the first or second
harmonic of the rf frequency. Magnetic resonance spectra were
recorded by sweeping the rf frequency and simultaneously recording
the in-phase and quadrature signals from the lock-in amplifier.  The
first harmonic and second harmonic signals were recorded in
sequential scans under identical conditions.

\section{DRAM theoretical model}
\label{sec:theory}

The double resonance alignment magnetometer model (DRAM model)
developed in~\cite{DRAMtheo} provides algebraic expressions for the
expected resonance signals.  In that model, the line shapes are
calculated following a three-step approach: creation of alignment by
optical pumping, magnetic resonance, and detection of the
oscillating steady state alignment.  In practice, the three steps
occur simultaneously and the approach is valid only if steady state
conditions are reached for the first two steps.  As explained
in~\cite{DRAMtheo}, the approach is thus valid only if the optical
pumping rate is negligible compared to the relaxation rates, i.e.,
for low light powers.  The model calculates the evolution of the
alignment multipole moments $m_{2,q}$ via a density matrix approach,
where the moments $m_{2,q}$ are defined with respect to a
quantization axis aligned with $\myvector{B}_0$ and relax with rates
$\sGamma{|q|}$. The third step probes the state of the $m_{2,q}$ and
results in detectable signals.

Only the most relevant equations needed for the analysis of the
presented measurements are reproduced here.  The magnetometer
signals which are modulated at the rf frequency $\omega$ and at its
second harmonic $2\omega$, can be written as
\begin{subequations}
\label{eq:signals}
\begin{eqnarray}
  S_{\omega }(t)&\!\!\!=
  \phantom{_2} \alignment\, h_{\omega}(\gamma) \left[ \right.&
    \!\!\! \phantom{-}D_{\omega }\,\cos \left(\omega t-\alpha\right)\nonumber \\
  && \!\!\! \left. -  A_{\omega }\,\sin \left(\omega
  t-\alpha\right)\right]
  \, ,
\label{subeq:somega}\\
  S_{2\omega }(t)&\!\!\!= \alignment\, h_{2\omega }(\gamma )\left[\right. &
  \!\!\!           -A_{2\omega }\,\cos \left(2\omega t-2\alpha\right)\nonumber \\
  && \!\!\! \left. -D_{2\omega }\,\sin \left(2\omega
  t-2\alpha\right)\right]
  \,, \quad \phantom{,}
  \label{subeq:s2omega}
\end{eqnarray}
\end{subequations}
where $\alignment$ is the alignment produced by the linearly
polarized laser light in the first step of the model.  The phase of
the rf field with respect to $\myvector{k}\times\myvector{B}_0$ is
$\alpha$ (cf.~Fig.~\ref{fig:geometry}).  The angular dependencies of
the first and second harmonic signals are given by
%
\begin{subequations}
\label{eq:angdep}
\begin{eqnarray}
h_{\omega }(\gamma )
&=&\frac{3}{16}\left( 2\sin
2\gamma +3\sin 4\gamma \right)\,,
\label{subeq:hgamma1} \\
h_{2\omega }(\gamma )
&=&\frac{3}{32}(1-4\cos 2\gamma +3\cos 4\gamma )\,,
\label{subeq:hgamma2}
\end{eqnarray}
\end{subequations}
where we have chosen an alternative representation of the
expressions given in~\cite{DRAMtheo}.
%
The first and second harmonic signals have both absorptive,
$A_{\omega}$, $A_{2\omega }$, and dispersive, $D_{\omega}$,
$D_{2\omega }$, components in their line shapes, given by
\begin{subequations}
\label{eq:lineshapes3gamma}
\begin{eqnarray}
D_{\omega}& = &\frac{\delta\, \sGamma{0}\omega_1
                     (\sGamma{2}^2+4 \delta^2-2 \omega_1^2)}{Z}
\, ,
\label{subeq:3gd1}\\
A_{\omega}& = &\frac{\phantom{\delta\,}\sGamma{0}\omega_1
                    \left[(\sGamma{2}^2 + 4 \delta^2) \sGamma{1}
                         + \sGamma{2} \omega_1^2 \right]} {Z}
\, ,
\label{subeq:3ga1}\\
D_{2\omega}& = &\frac{\delta\, \sGamma{0} \omega_1^2
                      (2\sGamma{1}+\sGamma{2})}{Z}
\, ,
\label{subeq:3gd2}\\
A_{2\omega}& = & \frac{\phantom{\delta\,}\sGamma{0}
                       \omega_1^2 (\sGamma{1}\sGamma{2} - 2\delta^2
                       +\omega_1^2)}{Z}
\, , \label{subeq:3ga2}
\end{eqnarray}
\end{subequations}
with
\begin{eqnarray}
Z&=& \sGamma{0} \left(\sGamma{1}^2+\delta^2\right) \left(\sGamma{2}^2
     + 4\delta^2 \right) \nonumber \\
 & &{}+\left[\sGamma{1} \sGamma{2} \left(2\sGamma{0} + 3\sGamma{2}\right)
     - 4 \delta ^2 \left(\sGamma{0} - 3\sGamma{1} \right) \right] \omega_1^2
      \nonumber \\
 & &{}+ \left(\sGamma{0} + 3\sGamma{2}\right) \omega_1^4 \,.
\label{eq:denom3gamma}
\end{eqnarray}

In Eqs.~(\ref{eq:lineshapes3gamma}) and (\ref{eq:denom3gamma}),
$\omega_{1}=\gamma_{F}B_{1}$ is the Rabi frequency of the rf field
where $\gamma_{F}$ is the Land\'e g-factor of the ground state
hyperfine level $F$.
The detuning, $\delta=\omega-\omega_{0}$, is the difference between
the radio frequency $\omega$ and the Larmor frequency
$\omega_{0}=\gamma_{F}B_{0}$.

\section{Analysis of the resonance signals}
\label{sec:signal}

\begin{figure}[t]
\centerline{\includegraphics[width=\grscale\textwidth]{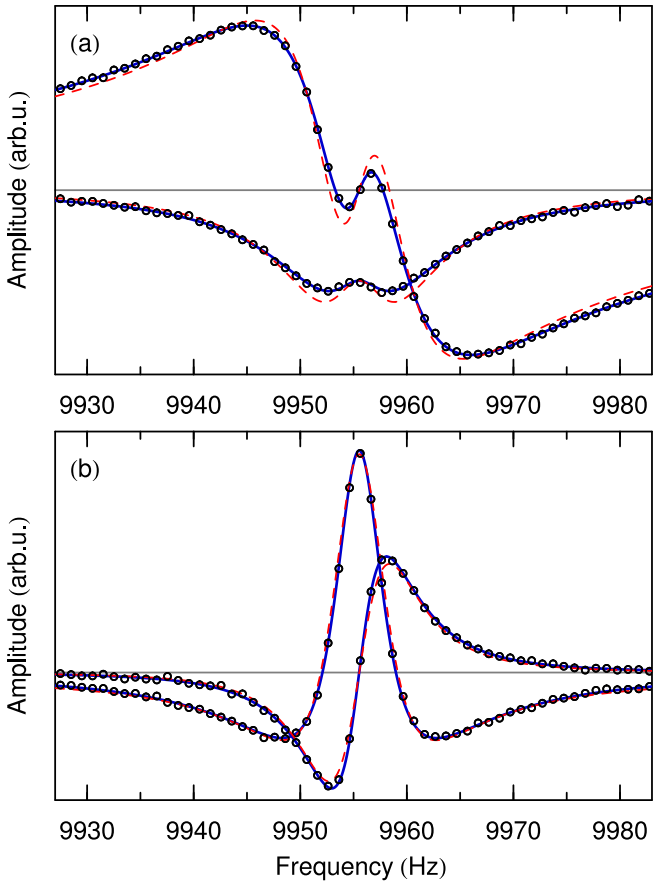}}
\caption{(Color online) Measurements (circles) of the in-phase and
quadrature magnetic resonance signals detected as amplitude
modulations of the transmitted light power. (a) First harmonic
signals. (b) Second harmonic signals. The statistical uncertainty on
each data point is smaller than the symbol size. The solid lines
(blue) are fits of the theoretical line shapes given by
Eqs.~(\ref{eq:lineshapes3gamma}) with three independent relaxation
rates. The dashed lines (red) are fits of the theoretical line
shapes given by Eqs.~(\ref{eq:lineshapes3gamma}) with only one
independent relaxation rate
$\Gamma\equiv\sGamma{0}=\sGamma{1}=\sGamma{2}$. The details of the
experimental conditions and fit results are in the text.}
\label{fig:resfit}
\end{figure}

%
Typical measurements of the in-phase and quadrature spectra of the
first and second harmonic signals are shown in
Fig.~\ref{fig:resfit}. Each pair of curves was recorded in a single
scan, with a lock-in time constant of 10~ms, and a sweep rate of
1~Hz/s.  Note that the same scan parameters were used for all other
magnetic resonance spectra presented in this paper.  The data of
Fig.~\ref{fig:resfit} were obtained with a laser light power of
$0.1\:\mu\mathrm{W}$ (approximate Gaussian profile, $1/e^2$ full
width of 3.7~mm horizontally and 3.2~mm vertically), an rf field
Rabi frequency of $\omega_{1}/(2\pi)=3.8~\Hz$, and an angle between
the linear polarization vector and the static magnetic field of
$\gamma=75 \deg$.

%
In practice, lock-in detection of the signals given by
Eqs.~(\ref{eq:signals}) with respect to the rf~frequency $\omega$
adds a phase \liaphase, selectable in the lock-in amplifier, and a
small pick-up signal $p_{(1,2)(A,D)}$ (smaller than 1\% of the
signal at maximum) to each of the line shapes in
Eqs.~(\ref{eq:lineshapes3gamma}).  The pick-up is due to direct
inductive coupling between the rf~field coils and the photodiode
readout wires: as expected, it varied with rf power but not with
$\delta$.  Due to \liaphase{}, the in-phase and quadrature spectra
are, in general, a mixture of dispersive and absorptive line shapes.
Indeed, after demodulation of the signal given by
Eq.~(\ref{subeq:somega}), we obtain the following expressions which
were used to fit the in-phase and quadrature spectra
\begin{subequations}
\label{eq:demod_one}
\begin{eqnarray}
I_\omega(\delta)\!
   &\! = g_{\omega} \alignment h_{\omega}(\gamma) &
    \kern-1ex
    \left[ \phantom{+}(D_{\omega}+p_{1D})\cos{(\alpha\!+\!\liaphase)}
    \right. \nonumber  \\
   &   & \kern-1ex \left.
           \, +(A_{\omega}+p_{1A})\sin{(\alpha\!+\!\liaphase)}
         \right]
\label{subeq:demod_ip} \\
Q_\omega(\delta)\!
   &\! = g_{\omega} \alignment h_{\omega}(\gamma) &
    \kern-1ex
    \left[ \phantom{+}(A_{\omega}+p_{1A})\cos{(\alpha\!+\!\liaphase)}
    \right.\nonumber  \\
   &   & \kern-1ex \left.
           \, -(D_{\omega}+p_{1D})\sin{(\alpha\!+\!\liaphase)}
         \right]
\label{subeq:demod_qu}
\end{eqnarray}
\end{subequations}
A similar mix of $A_{2\omega}$ and $D_{2\omega}$ was used for the
second harmonic signal given by Eq.~(\ref{subeq:s2omega}).  Due to
frequency dependent phase shifts in the signal treatment
electronics, both the effective \liaphase{} and the pickup terms
($p_{1A}$, $p_{1D}$, $p_{2A}$, $p_{2D}$) were different for the
$\omega$ and $2\omega$ signals.  Global detection factors,
$g_{\omega}$ and $g_{2\omega}$, were also required to model
amplifier dependent gains and light power dependent effects other
than the creation of $\alignment$. In total, nine free parameters
were needed to fit an in-phase plus quadrature signal pair, while
only thirteen were needed to fit all four signals simultaneously
since the physics parameters ($\omega_0$, $\omega_1$, \sGamma{0},
\sGamma{1}, and \sGamma{2}) were the same for both harmonics.  All
results for those parameters reported in this paper were taken, when
possible, from simultaneous four-spectra fits.

Leaving \dphase$=(\alpha+\liaphase)$ free during the fits to the
data had the advantage of simplifying the experimental procedure and
gave access to the phase information, which was used to adjust the
direction of the static magnetic field.  Indeed, for the
measurements testing the angular dependence of the resonance signals
(see \S\ref{sec:angulardep}), it is important that the static
magnetic field is exactly perpendicular to the laser beam wave
vector, as shown in Fig.~\ref{fig:geometry}.  This is the case when
the phase mismatch \dphase{} obtained from the fitting procedure is
independent of the angle $\gamma$ between the polarization and the
static magnetic field.

%
In order to check the validity of the model developed
in~\cite{DRAMtheo}, the theoretical line shapes were fitted to the
experimental data: the resulting curves are plotted with the
experimental data in Fig.~\ref{fig:resfit}.  The solid lines were
obtained using Eqs.~(\ref{eq:lineshapes3gamma}) with three
independent relaxation rates: $\sGamma{0}$ for the relaxation of
populations, $\sGamma{1}$ for the relaxation of $\Delta M=\pm1$
coherences, and $\sGamma{2}$ for the relaxation of $\Delta M=\pm2$
coherences.  As a result of the fit we obtained
$\sGamma{0}/(2\pi)=1.64(2)\:\Hz$, $\sGamma{1}/(2\pi)=2.93(2)\:\Hz$
and $\sGamma{2}/(2\pi)=3.08(2)\:\Hz$.

Note that the longitudinal relaxation rate $\sGamma{0}$ is quite
different from the transverse relaxation rates $\sGamma{1,2}$,
demonstrating that it is not possible to obtain a good fit with a
simpler model with only one single relaxation rate
$\Gamma\equiv\sGamma{0}=\sGamma{1}=\sGamma{2}$.  In that case the
line shape functions have a relatively simple algebraic form which
is given in~\cite{DRAMtheo}.  A fit of the simplified model is shown
as dashed curves in Fig.~\ref{fig:resfit} and yielded a
$\chi^2$-minimizing value of $\Gamma/(2\pi)=2.80(4)\:\Hz$. As
expected, the fit quality is poor compared to the model with three
independent relaxation rates.

Since the values of the transverse rates $\sGamma{1}$ and
$\sGamma{2}$ differ by less than 5\%, it is possible to use a model
with two relaxation rates, $\sGamma{L}\equiv\sGamma{0}$ and
$\sGamma{T}\equiv\sGamma{1}=\sGamma{2}$, without significantly
degrading the fit quality.  Curves representing a fit with only two
independent relaxation rates ($\sGamma{L}, \sGamma{T}$) are not
shown in Fig.~\ref{fig:resfit} because they are indistinguishable
from the solid lines obtained with three independent relaxation
rates.  However, we found that the ratio of $\sGamma{1}$ to
$\sGamma{2}$ depends on the angle $\gamma$, and that it changes with
laser power.  Therefore, the model with three relaxation rates will
be used for the rest of this work.

%
\begin{figure}[t]
\centerline{\includegraphics[width=\grscale\textwidth]{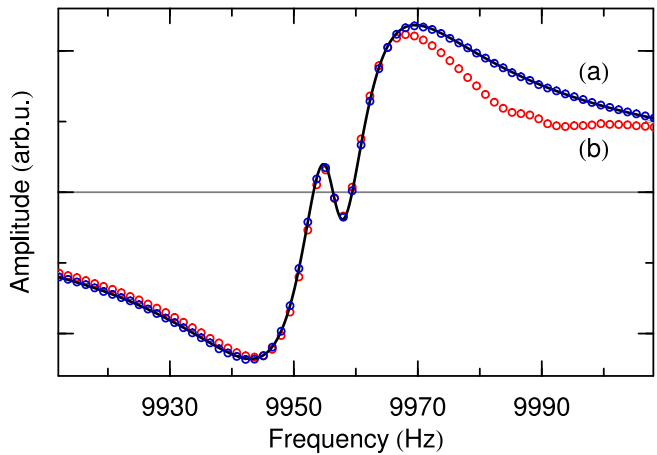}}
\caption{(Color online) Dispersive component of the magnetic
resonance signal oscillating at frequency $\omega$ measured under
the following conditions: the angle between the linear polarization
vector and the static field was $\gamma=25\deg$, the laser light
power was $0.1\:\mu\mathrm{W}$, and the rf field Rabi frequency was
$\omega_{1}/(2\pi)=4.8~\Hz$. Graph (a) (blue) was measured with a
rotating rf field and graph (b) (red) with an oscillating rf field.
The solid line is a fit of the theoretical line shape to the curve
obtained with the rf rotating field. The slight difference in
linewidth  suggests that the Rabi frequency in the two measurements
was not perfectly equal although we increased the amplitude of the
oscillating rf field by a factor 2.} \label{fig:deformation}
\end{figure}

Note that the use of a rotating rf field is crucial for the quality
of the fits. Magnetic resonance experiments, and in particular
optically pumped magnetometers, usually use a linearly polarized rf
field.  The implementation of such a field is technically less
demanding and its use is justified, in high-Q oscillators with a
single resonance frequency, by the rotating wave approximation by
virtue of which the precessing spins interact only with the
circularly polarized component of the rf field that co-rotates with
the spins.

When using an oscillating rf field one obtains the spectrum (b)
shown in Fig.~\ref{fig:deformation}, while the use of a rotating rf
field yields the spectrum (a). Spectrum (b) shows a second weak
resonance located approximately 33~Hz above the main resonance. That
difference frequency corresponds to the difference of the Larmor
frequencies of the $F_g=4$ and $F_g=3$ ground state hyperfine levels
due to the nuclear magnetic moment term in the expression for the
gyromagnetic ratio $g_F$.

The structure can be explained as follows: Optical pumping aligns
both the $F_g=3$ and the $F_g=4$ ground states and magnetic
resonance transitions can be driven within these two states.  The
resonance in the $F_g=3$ state is not detected directly since the
laser frequency is adjusted to the $F_g=4 \rightarrow F_e=3$
transition. However, spin exchange collisions can transfer the
alignment from the $F_g=3$ to the $F_g=4$ state where it can be
detected. In principle, this opens a way for the experimental study
of alignment transfer collisions. In our experiment, the use of a
rotating rf field allowed us to record spectra without contamination
from the $F_g=3$ state. This is due to the $g_F$-factors of both
states having opposite signs, so that they precess in opposite
directions, and the rotating field thus co-rotates with one species
only, i.e., the $F_g=4$ atoms.

\section{Angular dependence}
\label{sec:angulardep}

\begin{figure}
\centerline{\includegraphics[width=\grscale\textwidth]{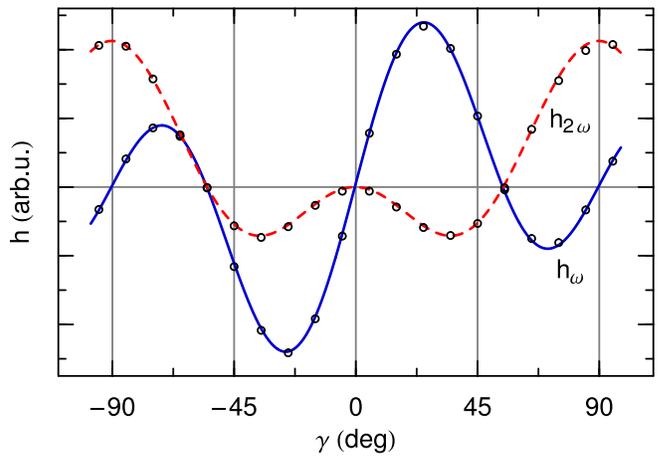}}
\caption{(Color online) Angular dependence of the amplitude of the
first (blue, solid line) and second (red, dashed line) harmonic
signals on the angle $\gamma$ between the light polarization
$\myvector{\epsilon}$ and the magnetic offset field
$\myvector{B}_{0}$.  Each experimental data point is the result of a
measurement and fit of the magnetic resonance spectra as explained
in \S\protect\ref{sec:signal}.  The measurements were made using a
light power of $0.07\:\mu\mathrm{W}$ and a rf field Rabi frequency
of $2.4~\Hz$. The lines were found by fitting the $h_{\omega }$ and
$h_{2\omega}$ functions given by Eqs.~(\protect\ref{eq:angdep}) to
the experimental data.} \label{fig:hom2om}
\end{figure}

%
Our theoretical DRAM model gives algebraic results which are valid
for arbitrary geometries.  To check the angular dependencies
predicted by the model we measured and fit the magnetic resonance
spectra for different values of $\gamma$, the angle between the
linear polarization vector and the static magnetic field. All fit
qualities were comparable to that of Fig.~\ref{fig:resfit}. From the
fits, the amplitudes of the first and second harmonic signals were
determined and the results are plotted as circles in
Fig.~\ref{fig:hom2om}.  The solid and dashed lines represent the
theoretical angular dependencies $h_{\omega}$ and $h_{2\omega}$
given by Eqs.~(\ref{eq:angdep}).  The agreement is very good.

%
A crucial point for these measurements was the quality of the linear
polarization, a parameter found to be particularly important for
small angles.  Indeed, both the first and second harmonic signals
should disappear when $\gamma$ is very small, and therefore any
small contamination of circular polarization produces a visible
parasitic resonance signal.  For that reason we used a high quality
Glan-prism to produce linear polarization with a degree of circular
polarization (DOCP) on the order of 0.1\%.  However, the
imperfection of the half-wave plate used to rotate the polarization
vector somewhat degraded the quality of the polarization.  In spite
of that, a~DOCP of less than 1\% could be maintained over the range
of angles $\gamma$ investigated, which proved to be sufficient for
the measurements.  Careful inspection of the angular dependence of
the second harmonic signal ($h_{2\omega}$ in Fig.~\ref{fig:hom2om})
reveals that all experimental points near $\gamma=0$ lie below the
theoretical curve. This is a typical effect from circular
polarization contamination, as verified by intentionally increasing
the DOCP, e.g., by using a polarizer of poorer quality.

\section{Dependence on the rf power}
\label{sec:rfpow}

\begin{figure}
\centerline{\includegraphics[width=\grscale\textwidth]{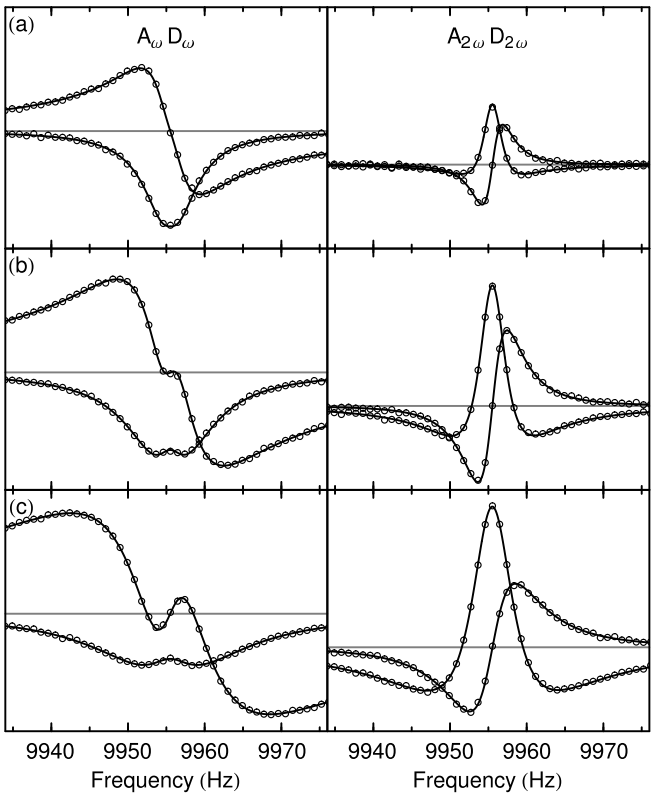}}
\caption{Resonance spectra measured with an angle $\gamma=75\deg$
and for different Rabi frequencies of the rf field: (a)
$\omega_{1}/(2\pi)=0.95~\Hz$, (b) $\omega_{1}/(2\pi)=2.4~\Hz$, (c)
$\omega_{1}/(2\pi)=4.8~\Hz$. The left column shows in-phase and
quadrature components of the first harmonic signal, and the right
column shows in-phase and quadrature components of the second
harmonic signal.  For these measurements, the light power was
$0.1\:\mu\mathrm{W}$.} \label{fig:resUrf}
\end{figure}

The DRAM theoretical model gives algebraic results which are valid
for arbitrary rf intensities. Its predictions were checked by
measuring and fitting magnetic resonance spectra for selected values
of the Rabi frequency of the rf field.  Three sets of illustrative
spectra are presented in Fig.~\ref{fig:resUrf} together with the
respective fits. The resonance curves were measured at a fixed angle
$\gamma=75\deg$ and for $\omega_{1}/(2\pi)=0.95~\Hz$, $2.4~\Hz$, and
$4.8~\Hz$.  The quality of the fits does not depend on the rf
intensity.  Moreover, all twelve curves were fitted with the same
unique set of physics parameters. In the presented data, the common
values for relaxation rates are $\sGamma{0}/(2\pi)=1.64(1)\:\Hz$,
$\sGamma{1}/(2\pi)=2.98(1)\:\Hz$ and
$\sGamma{2}/(2\pi)=3.05(1)\:\Hz$.  This shows that the evolution of
the resonance spectra with rf power is well described by the DRAM
theoretical model.

In particular, note the appearance of the narrow spectral feature in
the first harmonic signal as the rf power is increased.  This
feature emerges as predicted by the DRAM theoretical model for
$\omega_{1}>\sGamma{2}/\sqrt{2}$. As discussed in~\cite{DRAMtheo},
it can be explained as a product of the creation of a $\Delta M=2$
coherence by a second order interaction with the rf field, followed
by the evolution of that coherence in the offset field, and
back-transfer to a $\Delta M=1$ coherence by an additional
interaction with the rf field, which can then be detected at the
first harmonic frequency $\omega$. In our measurement,
$\sGamma{2}/\sqrt{2} = 2\pi \times 2.16\:\invs$ and we indeed
observe this narrow spectral feature when $\omega_{1}/(2\pi)$
becomes larger than $2.16\:\Hz$. Note that this feature allows an
easy calibration of the rf voltage applied to the coils in terms of
the resulting Rabi frequency $\omega_{1}$.

\begin{figure}
\centerline{\includegraphics[width=\grscale\textwidth]{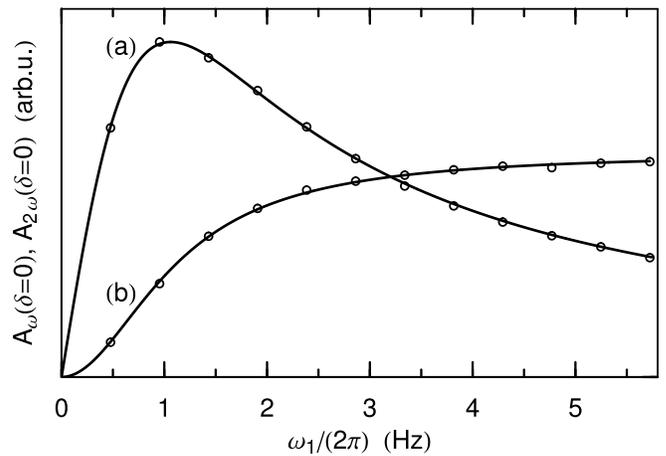}}
\caption{Saturation of the resonant absorptive signals $A_{\omega}$
and $A_{2\omega}$ as a function of the Rabi frequency $\omega
_{1}/(2\pi)$ of the rf field. (a) On-resonance amplitude of the
first harmonic signal measured for an angle $\gamma=25.5\deg$. (b)
On-resonance amplitude of the second harmonic signal measured for an
angle $\gamma=90\deg$. For these measurements, the light power was
fixed to $0.1\:\mu\mathrm{W}$. The solid lines are fits of the DRAM
theoretical model [Eqs.~(\ref{eq:rfsat})] to the experimental data.}
\label{fig:Amax}
\end{figure}

%
In view of potential applications in magnetometry the angles
$\gamma$ which maximize the resonance signals are of particular
interest. Theory predicts $\gamma=25.5\deg$ for the maximum of the
first harmonic and $\gamma=90\deg$ for the second harmonic signal
(cf.~Fig.~\ref{fig:hom2om}).  We studied the saturation of the
absorptive component of the resonance spectra at those angles by
changing the amplitude of the rf~field and determining the
on-resonance amplitudes, i.e., $A_{\omega}(\delta\!=\!0)$ for
$\gamma=25.5\deg$ and $A_{2\omega}(\delta\!=\!0)$ for $\gamma=90\deg$.
The results are presented in Fig.~\ref{fig:Amax}, together with fits
of the DRAM theoretical model functions given by:
\begin{subequations}
\label{eq:rfsat}
\begin{eqnarray}
A_\omega(\delta=0)& = &\frac{\sGamma{0}\sGamma{2}\omega_1}
                     {\sGamma{0}\sGamma{1}\sGamma{2}+(\sGamma{0}+3\sGamma{2})\omega_1^2}
\, ,
\label{subeq:rfsat1}\\
A_{2\omega}(\delta=0)& = &\frac{\sGamma{0}\omega_1^2}
                     {\sGamma{0}\sGamma{1}\sGamma{2}+(\sGamma{0}+3\sGamma{2})\omega_1^2}
\, .
\label{subeq:rfsat2}
\end{eqnarray}
\end{subequations}
The absorptive component of the first harmonic signal goes through a
maximum at $\omega_{1}/(2\pi)=1.05\:\Hz$ and then decays to zero
when the rf power is further increased, while the second harmonic
signal saturates at a constant value when the rf amplitude tends to
infinity. The excellent agreement of the model function with the
experimental data proves the validity of the DRAM model for
arbitrary rf intensities.

\section{Dependence on light power: limits of validity of the model}
\label{sec:lightpow}

%
\begin{figure}
\centerline{\includegraphics[width=\grscale\textwidth]{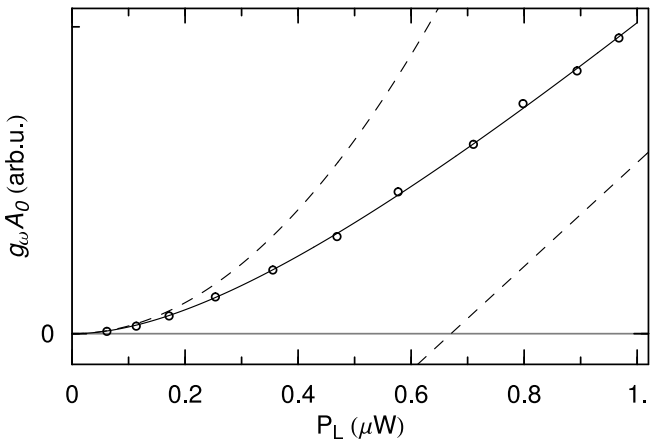}}
\caption{Global amplitude factor $g_{\omega}\alignment$ from
Eq.~(\protect\ref{eq:demod_one}) of the first harmonic resonance
signal as a function of the laser power $P_L$ measured for
$\gamma=25.5\deg$. Each experimental data point is the result of a
measurement and a fit of the magnetic resonance spectra as explained
in \S\protect\ref{sec:signal}. For these measurements, the Rabi
frequency of the rf field was fixed at $\omega_{1}/(2\pi)=3.8~\Hz$.
The solid line is a fit of the function given by
Eq.~(\protect\ref{eq:Asat}) to the experimental data. The dashed
lines represent the quadratic parts of the fit function and the
linear asymptote to Eq.~(\protect\ref{eq:Asat}) which has a zero
crossing at the saturation power
$P^{\mathrm{sat}}_{\omega}=0.67(3)\:\mu\mbox{W}$.} \label{fig:PL}
\end{figure}

As mentioned before, the validity of the three step model is limited
to low laser powers. In that regime, one expects a quadratic
dependence of the signals on laser power $P_L$.  This is due to the
fact that (in lowest order) the alignment produced in step one is
proportional to $P_L$ and that the detection process (measurement of
light absorption/transmission) is also proportional to $P_L$.  The
limits of the model's validity are thus expected to show up as a
deviation from a quadratic power dependence when increasing the
laser power. To check this point, we have measured the magnetic
resonance spectra for different values of the laser power. For each
value of $P_L$, the first and second harmonic signals were fitted
using the procedure described in \S\ref{sec:signal}, and the global
amplitude factors $g_{\omega}\alignment$ and $g_{2\omega}\alignment$
of Eqs.~(\ref{eq:demod_one}) were determined (remember that
$\alignment$ is the alignment initially created in step one of the
model, and the $g_{\omega}$, $g_{2\omega}$ factors are experimental
detection amplitude gains). The results are plotted in
Fig.~\ref{fig:PL} for $g_{\omega}\alignment$ measured at an angle
$\gamma=25.5\deg$. The graph for $g_{2\omega}\alignment$ is very
similar to that of $g_{\omega}\alignment$ and is therefore omitted.
Moreover, the following discussion of the dependence on laser power
of the first harmonic signals is also valid for the second harmonic.

%
\begin{figure}
\centerline{\includegraphics[width=\grscale\textwidth]{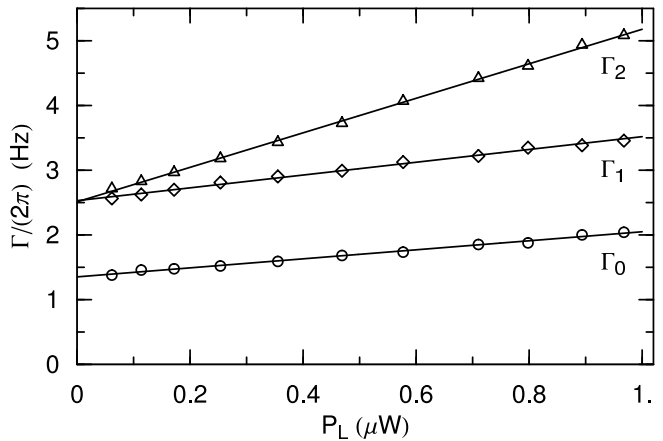}}
\caption{Relaxation rates as a function of the laser power $P_L$
measured for $\gamma=25.5\deg$. Each triplet
$(\Gamma1,\Gamma2,\Gamma3)$ is the result of a measurement and a fit
of the magnetic resonance spectra as explained in
\S\protect\ref{sec:signal}, starting from the same experimental data
which were used for Fig.~\protect\ref{fig:PL}. The circles represent
$\sGamma{0}/(2\pi)$, the squares $\sGamma{1}/(2\pi)$ and the
triangles $\sGamma{2}/(2\pi)$. The solid lines are linear fits to
the experimental data.} \label{fig:relax}
\end{figure}

As expected, we observe in Fig.~\ref{fig:PL} that the resonance
signals increase quadratically with $P_L$ for low light power only.
We found that the power dependence is well represented by the
empirical function
\begin{equation}
\label{eq:Asat} g_{\omega}\alignment = C_{\omega} \frac{G^2}{1+G}
\end{equation}
where $C_{\omega}$ is a constant and
$G=P_L/P^{\mathrm{sat}}_{\omega}$ is a saturation parameter which
measures the applied light power normalized to the saturation power
$P^{\mathrm{sat}}_{\omega}$.  As can be seen in Fig.~\ref{fig:PL},
the empirical dependence of Eq.~(\ref{eq:Asat}) gives an excellent
fit to the data.  The fit of the data in Fig.~\ref{fig:PL} yields
$P^{\mathrm{sat}}_{\omega}=0.67(3)\:\mu\mbox{W}$.

This is a quite astonishing but very satisfactory result.  It is
astonishing because the power dependence cannot be reproduced in a
simple way by an extension of the model in~\cite{DRAMtheo}.
Equation~(\ref{eq:Asat}) can be interpreted in terms of two factors:
the first one, $G/(1+G)$, describing the saturation of the initial
alignment creation and the second one, $G$, describing the probing
of the steady state alignment.  For a closed $J=1/2\rightarrow
J=1/2$ transition pumped by circularly polarized radiation (DROM)
one can prove the exact validity of that dependence for arbitrary
light powers $P_L$. However, for transitions $J\rightarrow J'$
between states of arbitrary angular momenta, linearly polarized
optical pumping produces not only a ground state alignment but also
higher order multipole moments and each interaction with the light
transfers populations and coherences back and forth between these
multipole moments.  In this sense it is astonishing that the probing
process which detects only alignment components leads to the simple
power dependence given by Eq.~(\ref{eq:Asat}).  The result is
satisfactory because it gives an algebraic expression for the DRAM
signals even for light powers which exceed the anticipated validity
limit of the model.

In Fig.~\ref{fig:relax} we have plotted the relaxation rates
corresponding to the measurements of Fig.~\ref{fig:PL}. They
increase linearly with laser power, as expected, to first order,
when taking into account the depolarization of light interactions
(see \cite{DRAMtheo}). For low light power, we see that
$\sGamma{1}\approx\sGamma{2}$ and therefore a model with two
relaxation rates ($\sGamma{L}\equiv\sGamma{0}$ and
$\sGamma{T}\equiv\sGamma{1}=\sGamma{2}$) is sufficient. However, as
soon as $P_L$ becomes non-negligible compared to
$P^{\mathrm{sat}}_{\omega}$, a model with three independent
relaxation rates becomes necessary. The values of the relaxation
rates extrapolated to zero laser power are
$\sGamma{0}/(2\pi)=1.35(1)\:\Hz$, $\sGamma{1}/(2\pi)=2.53(1)\:\Hz$
and $\sGamma{2}/(2\pi)=2.51(1)\:\Hz$. These values are cell and
temperature dependent, but do not depend on the angle $\gamma$
between the light polarization and the static magnetic field.
However, we observed that the rate at which the relaxation rates
increase with laser power depend on $\gamma$.

\section{Conclusion}
\label{sec:conclusion}

In this experimental work, we have verified the validity of the very
general DRAM theoretical model for double resonance alignment
magnetometers developed in~\cite{DRAMtheo}.  We have shown that it
is valid for any geometry (i.e., any relative directions of the
laser beam, the light polarization vector, and the static magnetic
field) and for any rf power, as long as the laser power is kept
small. Moreover, we have investigated the role of laser power to
determine the domain of parameters for which the DRAM model is
valid. The measurement of the laser power dependence allowed us to
extend the DRAM model with an empirical algebraic formula to light
powers which exceed the expected validity limit of the model.
Finally, the data analysis revealed that three relaxation rates are
necessary to fit the DRAM model to the experimental data, which, in
principle, opens a way to investigate in detail relaxation
mechanisms in aligned alkali vapors.

\begin{acknowledgments}
This work was supported by grants from the Swiss National Science
Foundation (Nr.~205321--105597), from the Swiss Innovation
Promotion Agency, CTI (Nr.~8057.1 LSPP--LS) and from the Swiss
Heart Foundation.
\end{acknowledgments}


\end{document}